\documentclass[a4paper,11pt]{article}

\usepackage{jcappub} 
\usepackage{caption}
\usepackage{subcaption}
\usepackage{wrapfig}
\usepackage{aas_macros}
\usepackage{mathbbol}
\usepackage{slashed}
\usepackage{xcolor}
\usepackage{enumitem}
\usepackage{hyperref}
\usepackage{orcidlink}

\title{\boldmath Probing multi-state dark matter via optical absorption lines in DESI spectra}
\author[a]{Anoma Ganguly,\orcidlink{0000-0002-5915-4245}}
\author[b]{Shadab Alam \orcidlink{0000-0002-3757-6359}}

\affiliation[a]{Department of Astronomy, University of Arizona, Tucson, AZ 85721, USA}
\affiliation[b]{Department of Theoretical Physics, Tata Institute of Fundamental Research,\\Homi Bhabha Road, Mumbai 400005, India}

\emailAdd{anomaganguly@arizona.edu}
\abstract{We search for dark matter through absorption lines imprinted on the spectra of background sources whose sight lines pass through foreground dark matter halos. Such a line is a generic feature of multi-state dark matter models that couple to photons through higher-order electromagnetic moments. We focus on the minimal case of a two-state system with a magnetic dipole transition, in which the ground state absorbs a photon and makes a transition to the excited state. For transition energy $\Delta E \sim \mathcal{O}(\text{eV})$, the line falls in the optical band, within reach of spectroscopic surveys such as DESI. Stacking 145{,}087 spectra of DESI Early Data Release objects paired with foreground GAMA galaxy groups, we find a stacked optical depth consistent with zero and place $2\sigma$ upper limits on the magnetic dipole transition cross-section. The limits are strongest at $\Delta E \sim 3$--$5$\,eV for a few MeV dark matter mass in the parameter space that is inaccessible to both cosmological (BBN+CMB) and direct detection experiments. We forecast that the full DESI and the next-generation spectroscopic surveys will tighten these limits by more than an order of magnitude, establishing spectroscopic stacking as a competitive probe of electromagnetically interacting multi-state dark matter. 
}

\begin{document}
	{\tiny {\tiny}}	\maketitle
	\flushbottom
	
	\section{Introduction}
    The particle nature of dark matter remains one of the central open questions. While its gravitational influence is firmly established across cosmological and astrophysical scales, decades of searches for non-gravitational interactions have yet to yield a confirmed signal. This has motivated broad interest in candidates spanning a wide range of masses and couplings to the Standard Model, and in observational probes capable of probing such candidates.
    
    A well-motivated possibility is that dark matter is not a single-state particle but consists of multiple states that interact electromagnetically. While dark matter with non-zero electric charge is tightly constrained, the multi-state dark matter can be electrically neutral and yet couple to the photon through higher-order electromagnetic moments~\cite{2000PhLB..480..181P, 2004PhRvD..70h3501S, 2011PhLB..696...74B}. The radiative transitions between dark matter states can give rise to absorption or emission features, thus opening the possibility of detecting such models by looking for new signatures in the spectra of astrophysical sources.
    
    We consider a minimal scenario, a two-state system in which a \emph{transition} magnetic dipole connects a ground state to a nearly degenerate excited state. This magnetic dipole coupling makes the dark matter inherently inelastic, a structure previously invoked with $\mathcal{O}(100 \text{keV})$ mass splittings to reconcile direct detection anomalies~\cite{2010PhRvD..82l5011C}. We focus instead on a largely unexplored regime in which the resonant absorption $\chi + \gamma \to \chi^*$ excites the ground state and removes a photon of energy $\Delta E$ from the incident radiation. For $\Delta E \sim \mathcal{O}(\text{eV})$, this absorption falls in the optical band, placing a sharp, monochromatic dark matter signature directly within the wavelength coverage of spectroscopic galaxy surveys.
    
    In earlier work, we showed that such radiative dark matter transitions can imprint observable distortions on the cosmic microwave background and the global 21\,cm signal~\cite{Ganguly2024, Ganguly2025}. In this work, we search for the dark matter \emph{absorption} line against the light of optical background sources. An absorption signature offers several advantages over the corresponding emission signal: it is enhanced for a bright background source. It is sensitive to dark matter across different foreground halo masses and density profiles, rather than only to the densest emitting regions. Since the optical depth depends linearly on the dark matter column density along the sight line, the signal is only fairly insensitive to the detailed inner structure of the halo.
    
    The expected absorption from any single halo is far too weak to detect, so we adopt a stacking approach. By aligning many spectra in the rest frame of their respective foreground halos, the universal dark matter line accumulates coherently while uncorrelated noise and astrophysical contaminants average away. We construct source--absorber pairs from the Dark Energy Spectroscopic Instrument (DESI) Early Data Release~\cite{DESI_EDR2024} and foreground galaxy groups from the Galaxy And Mass Assembly (GAMA) survey~\cite{Robotham2011}.
    Stacking 145{,}087 source--absorber pairs, we find a stacked optical depth consistent with zero and place the first observational upper limits on the transition magnetic dipole cross-section in the optical regime. The constraints already reach into otherwise inaccessible parameter space, and the forecast for future surveys will tighten them by more than an order of magnitude.
    
    The remainder of this paper is organised as follows. Section~\ref{sec:dark line} introduces the dark matter model and derives the predicted absorption signature. Section~\ref{sec:data} describes the DESI and GAMA datasets. Section~\ref{sec:methodology} presents the pair-selection, stacking, and constraint-derivation pipeline. Section~\ref{sec:results} reports the resulting limits and forecasts the improvement with future surveys, and Section~\ref{sec:conclusion} presents the conclusions.

    \section{Theoretical model}

    \label{sec:dark line}

    \subsection{Dark matter model}
    We consider a minimal dark matter model comprising two electrically neutral states. The ground state is a pseudoscalar $\chi$ with mass $m_\chi$ and the excited state is a vector $\chi^*_\mu$ with mass $m_{\chi^*}$. This two-state structure can be compactly represented as a matrix multiplet,
    \begin{equation}\label{eq:multiplet}
        \mathcal{X} \;\equiv\; \chi^*_\mu\,\gamma^\mu \;-\; \chi\,\gamma^5 \,,
    \end{equation}
    The two states are separated by a small mass splitting $\Delta m \equiv m_{\chi^*} - m_\chi \ll m_\chi$, so that the radiative transition $\chi^* \to \chi + \gamma$ produces a photon with energy $\Delta E = \Delta m\, c^2$ far below the rest-mass scale of either state. Thus, the effective theory governing the radiative transition is non-relativistic. Following the standard  heavy quark effective theory 
    (HQET) procedure~\cite{Isgur:1989vq, Georgi:1990um, Wise:1992hn, Yan:1992gz, 
    1994NuPhB.412..181J, Mehen:2005hc, 1997PhR...281..145C}, we define velocity-labelled fields:
    \begin{equation}\label{eq:nrfield}
        \mathcal{X}_v \;\equiv\; \sqrt{2\,m_\chi}\; 
        e^{\,i\, m_\chi\,(v \cdot x)}\; P_+\, \mathcal{X} \,,
        \qquad\text{where}\qquad
        P_+ \;=\; \frac{1}{2}\bigl(1 + v_\mu\,\gamma^\mu\bigr) 
        \quad\text{and}\quad v^2 = 1 \,,
    \end{equation}
    where $v^\mu$ is the four-velocity of the dark matter particle. The projection operator $P_+$ removes the large-momentum modes, retaining only the small fluctuations with momenta $\ll m_\chi c$. A possible UV-complete model giving rise to this effective theory is developed in~\cite{Ganguly2024}.
     
    The radiative transition between the two states occurs via the magnetic dipole  operator~\cite{2000PhLB..480..181P},
    \begin{equation}\label{eq:M1operator}
        \mathcal{L}_{\mathcal{M}} \;=\; 
        -i\frac{\epsilon\, e}{4\,m_\chi}\;
        \mathrm{tr}\!\left(\bar{\mathcal{X}}_v\,\sigma^{\mu\nu}\,
        \mathcal{X}_v\,F_{\mu\nu}\right) = -\frac{i}{4}\mathcal{M}\;
        \mathrm{tr}\!\left(\bar{\mathcal{X}}_v\,\sigma^{\mu\nu}\,
        \mathcal{X}_v\,F_{\mu\nu}\right) ,
    \end{equation}
    where $F_{\mu\nu}$ is the electromagnetic field-strength tensor, 
    $\sigma^{\mu\nu} = \tfrac{i}{2}[\gamma^\mu,\gamma^\nu]$. To compare our results with the literature, we define the magnetic dipole cross-section as,
\begin{equation}\label{eq:sigmam}
    \sigma_\mathcal{M} = 2 \alpha \mathcal{M}^2 
    \,.
\end{equation}
     
    \medskip
    \noindent\textbf{Connection to magnetic dipole dark matter:}\quad
    The operator in Eq.~\eqref{eq:M1operator} generalizes the diagonal magnetic dipole dark matter framework~\cite{2004PhRvD..70h3501S, 2011PhLB..696...74B}, in which a single Dirac fermion couples to the photon via $-i/2\,\mathcal{M}\bar{\chi}\,\sigma^{\mu\nu}\chi\,F_{\mu\nu}$, to the case where the dipole connects two distinct mass eigenstates $\chi$ and $\chi^*$, making the interaction \emph{inelastic}. A similar off-diagonal magnetic dipole was introduced by~\cite{2010PhRvD..82l5011C} to explain direct detection anomalies with $\mathcal{O}(100\,\,\text{keV})$ splittings. In this work, we focus on the regime, $\Delta E \sim$ a few \text{eV}, where the radiative transition $\chi + \gamma \to \chi^*$ produces an absorption line in the optical or near-UV band, placing it within the wavelength coverage of large spectroscopic surveys such as DESI.

\subsection{Absorption signature}\label{sec:absorption}
 
The magnetic dipole moment operator in Eq.~\eqref{eq:M1operator} allows dark matter particles in the 
ground state $\chi$ to absorb photons at the transition wavelength 
$\lambda_* = hc/\Delta E$ and transition to the excited state $\chi_*$. A dark matter halo therefore acts as an absorber: when the line of sight (LoS) to a background source 
passes through a dark matter halo located at redshift $z_\mathrm{abs}$, the  absorption by dark matter particles imprints an absorption feature in 
the observed spectrum at the redshifted wavelength $\lambda=\lambda_*(1+z_\mathrm{abs})$. The resultant attenuation in the incident flux is characterised by the optical depth $\tau_\lambda$. For a LoS passing through a halo at an impact parameter $p$, the optical depth is given by~\cite{2002ApJ...579....1F},
\begin{equation}\label{eq:tau}
    \tau_\lambda(p) \;=\; 
    \frac{2\pi^3\,\mathcal{M}^2\,\lambda_*}{h\,c}
    \int_\text{LoS}
    \frac{\rho_\mathrm{abs}(r)}{m_\chi}\;
    \phi_\lambda(r,\, p)\;
    \left(
        \frac{1 - e^{-hc/(\lambda_*\,k_\mathrm{B}T_\text{ex}}}
             {1 + 3\,e^{hc/(\lambda_*\,k_\mathrm{B}T_\text{ex}}}
    \right) ds \,,
\end{equation}
where $s$ denotes the LoS distance to the halo, $r = \sqrt{p^2 + s^2}$ is the radial distance from the halo centre, $\rho_\mathrm{abs}$ is the absorber halo dark matter density profile, 
$T_\text{ex}$ is the excitation temperature governing the level population of dark matter particles in the two states. We assume that the two dark matter states are in equilibrium with the CMB photons $T_\mathrm{ex} = T_\mathrm{CMB}\,(1+z_\mathrm{abs})$. The factor in parentheses accounts for the population difference between the ground spin-0 state and the excited spin-1 state. The random motion of the dark matter particles in the halo gives rise to the Doppler broadening of the absorption line characterized by the line profile\footnote{We place the line at the cosmological redshift and neglect
the halo's peculiar line-of-sight velocity, which shifts the center by
$v_\mathrm{los}/c\sim10^{-3}$ ($v_\mathrm{los}$), sub-percent relative to the cosmological redshift.},
    \begin{align}
        \phi_\lambda(r, p) &= \frac{1}{\sqrt{\pi}\,
        \Delta\lambda_D(r)}\exp\!\left(-\frac{\bigl(\lambda 
        - \lambda_*\bigr)^2}{\Delta\lambda_D(r)^2}
        \right), \quad
        \text{where}\,\,\,
        \Delta\lambda_D(r) = \sqrt{\frac{2}{3}}\lambda_*\frac{\sigma_\mathrm{abs}}{c}\,,
        \label{eq:line_profile}
    \end{align}
where $\sigma_\mathrm{abs}$ denotes the halo 3D velocity dispersion  obtained using the fitting formula given in \cite{Ascasibar2004}.

With the optical depth parametrization given in Eq.~\eqref{eq:tau}, our dark matter model is fully specified by three free parameters: the dark matter mass $m_\chi$, the transition energy (wavelength) $\Delta E$ ($\lambda_*$), and the magnetic dipole moment $\mathcal{M}$.

\section{Data}
\label{sec:data}
Our search combines a catalogue of galaxy groups, which defines the
target population and supplies positions and redshifts, with the
publicly released DESI spectra in which we search for the absorption
signature predicted by our dark matter model. We describe these datasets below.

\subsection{The GAMA Galaxy Group Catalogue}
\label{sec:data:gama}

The group sample is drawn from the Galaxy And Mass Assembly (GAMA)
Galaxy Group Catalogue ($\mathrm{G^3C}$)~\cite{Robotham2011}, built on
the GAMA spectroscopic redshift survey~\cite{Driver2011} with input
photometry and target selection described
in~\cite{Baldry2010,Robotham2010tiling}. Groups in the
$\mathrm{G^3C}$ are identified with a friends-of-friends (FoF)
algorithm operating in projected separation and line-of-sight velocity,
with the linking lengths calibrated against mock GAMA lightcones
constructed from $\Lambda$CDM $N$-body simulations populated with
semi-analytic galaxies, so that the recovered group memberships and
derived properties are robust to interlopers~\cite{Robotham2011}. For
each group the catalogue provides a membership list, several
estimates of the group centre, a redshift, an intrinsic velocity
dispersion (computed with the gapper estimator), a characteristic
radius, and a dynamical mass estimate of the form
$M_{\mathrm{dyn}} \propto \sigma^2 R$~\cite{Robotham2011}. We use groups in G09, G12, G15 and G02 regions and select the ones with more than 3 members.

\subsection{DESI Spectra: the Early Data Release}
\label{sec:data:desi}

The spectra are taken from the Early Data Release (EDR) of the Dark
Energy Spectroscopic Instrument (DESI)~\cite{DESI_EDR2024}. DESI is a
multi-object fibre spectrograph mounted at the prime focus of the
4\,m Mayall telescope at Kitt Peak National Observatory, capable of
recording up to $5000$ spectra simultaneously over a
$\sim 8\,\mathrm{deg^2}$ field of view~\cite{DESI2016a,DESI2016b,DESI_Instrument2022}. The spectra span the wavelength range $3600$--$9800\,\text{\AA}$ across three arms, with
a resolving power that increases from $R\sim2000$ in the blue to
$R\sim5500$ in the near-infrared~\cite{DESI_Instrument2022}, which sets
the velocity resolution available to our line search.

The EDR comprises the spectra obtained during the five-month Survey
Validation (SV) campaign carried out between December 2020 and June
2021, prior to the DESI Main Survey~\cite{DESI_SV2024,DESI_EDR2024}.
Raw frames are reduced to wavelength- and flux-calibrated,
sky-subtracted one-dimensional spectra by the DESI spectroscopic
pipeline~\cite{Guy2023}, following the survey operations strategy
of~\cite{Schlafly2023}; spectral classifications and redshifts are
derived by the \textsc{Redrock} template-fitting code.\footnote{\url{https://github.com/desihub/redrock}}
Targets are selected from the DESI Legacy Imaging
Surveys~\cite{Dey2019}. The low-redshift galaxies that overlap the GAMA group sample fall predominantly within the DESI Bright Galaxy Survey
(BGS) target class~\cite{Hahn2023}.

\section{Methodology}\label{sec:methodology}
The dark matter absorption signature discussed in Sec.~\ref{sec:absorption} is far weaker than the noise of the spectrum of a single DESI object. Stacking many spectra allows the common absorption line (if any) in the rest-frame of the foreground halo to accumulate coherently, while uncorrelated noise averages down. 

We search for the dark matter absorption line by stacking the spectra of background DESI galaxies and quasars (\emph{sources}) whose lines of sight pass through foreground GAMA galaxy group halos (\emph{absorbers}). This section describes the three stages of the analysis: identifying source--absorber pairs (Sec.~\ref{sec:pairs}), constructing the stacked spectrum (Sec.~\ref{sec:stacking}), and deriving the constraints on the transition magnetic dipole moment $\mathcal{M}$ (Sec.~\ref{sec:constraints}).

\subsection{Source--absorber pair identification}
\label{sec:pairs}
We select GAMA groups with friends-of-friends multiplicity $N_\text{fof} > 3$ for reliable dynamical mass estimates. A source--absorber pair is retained if it satisfies two conditions:
\begin{itemize}
    \item \textbf{Angular proximity.} The angular separation $\theta$ between source and absorber centre satisfies,
    \begin{equation}\label{eq:theta100}
        \theta < \theta_{100} = \frac{R_{100}}{d_A(z_\text{abs})} \,,
    \end{equation}
    where $R_{100}$ is the radius enclosing a mean density of $100$ times the critical density $\rho_\text{crit}$ and $d_A(z_\text{abs})$ is the angular diameter 
    distance to the absorber. This samples the dense inner halo where the dark matter column density, and hence the optical depth, is largest.
 
    \item \textbf{Redshift separation.} The source must lie behind the 
    absorber, $z_\text{source} - z_\text{abs} > 0.02$. This buffer 
    corresponds to $ v/c \sim 10^{-2}$, which is greater than the typical velocity dispersion of massive halos. This removes the source galaxies that are members of the GAMA group. This exclusion is essential because the member galaxies sit at the absorber redshift, and their atomic and molecular lines would add coherently in the stack and overwhelm the much weaker dark matter feature.
\end{itemize}
    \begin{figure}[!htbp]
        \centering
        \includegraphics[width=\linewidth]{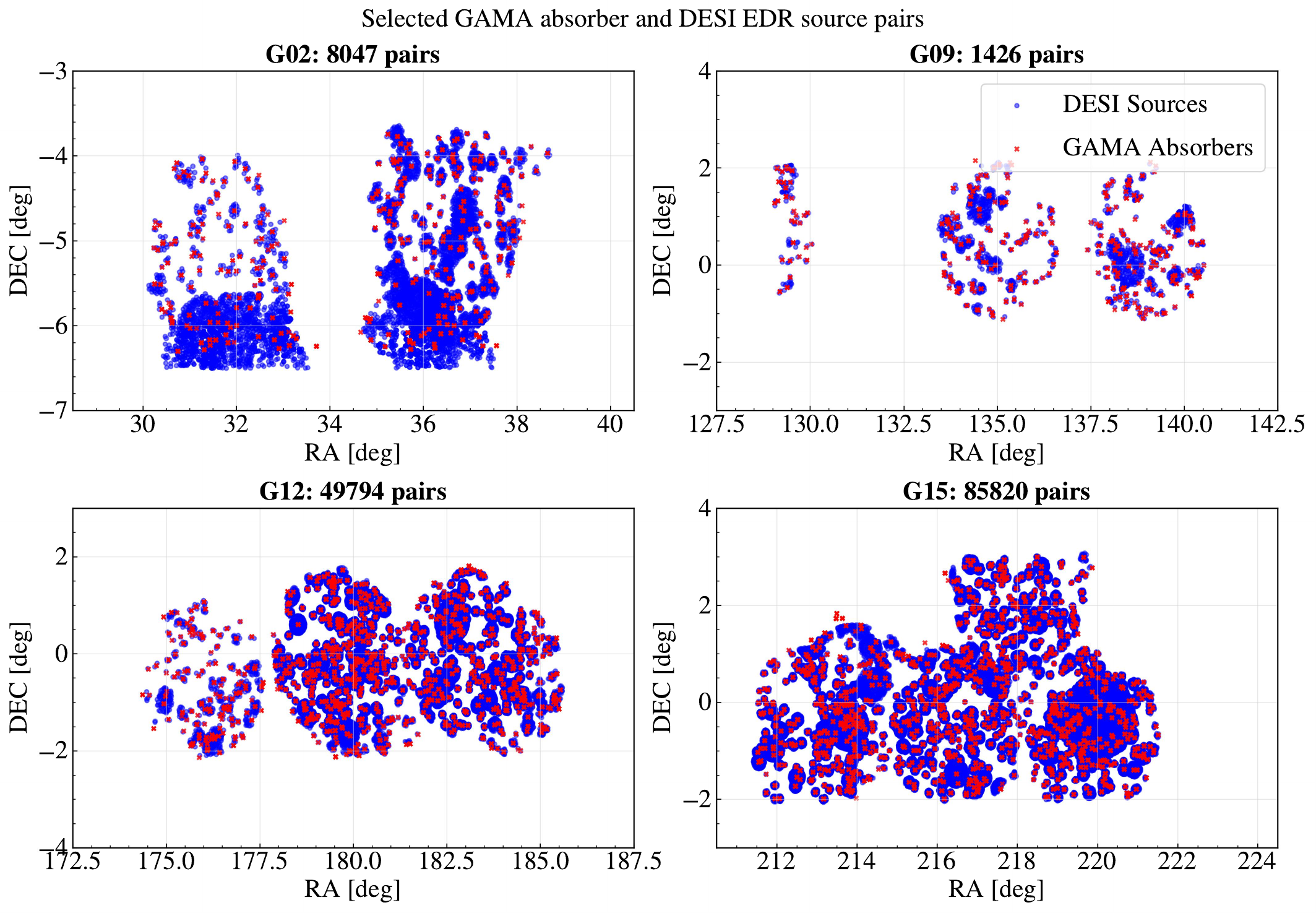}
        \caption{Sky distribution in equatorial coordinates of the DESI EDR background sources (blue dots) and GAMA foreground galaxy groups (red crosses) that satisfy the pair selection criteria in Sec.~\ref{sec:pairs}. Each panel corresponds to one of the four GAMA survey regions, with the total number of source--absorber pairs indicated. The combined sample yields 145,087 pairs across all four fields. 
        }
        \label{fig:selected_pairs}
    \end{figure}
Applying these criteria across the four GAMA fields yields a total of 145,087 source--absorber pairs, whose sky distribution is shown in Fig.~\ref{fig:selected_pairs}. The stacking procedure is robust against astrophysical contamination because the dense interstellar gas that produces atomic and molecular lines is confined to galaxy disks of $\sim\!20$\,kpc, far smaller than the typical size $R_{100} \sim\!\text{Mpc}$ of a galaxy group dark matter halo. Thus, a given sight line is roughly two orders of magnitude more likely to intercept dark matter than galactic gas. Second, only the lines correlated with the absorber stacks coherently. The intervening absorption lines from the Lyman-$\alpha$ forest or metal systems, when shifted to the absorber rest frame, average into a smooth, featureless background. 
    
\subsection{Spectral processing and stacking}\label{sec:stacking}
      \begin{figure}[!htbp]
        \centering
        \includegraphics[width=\linewidth]{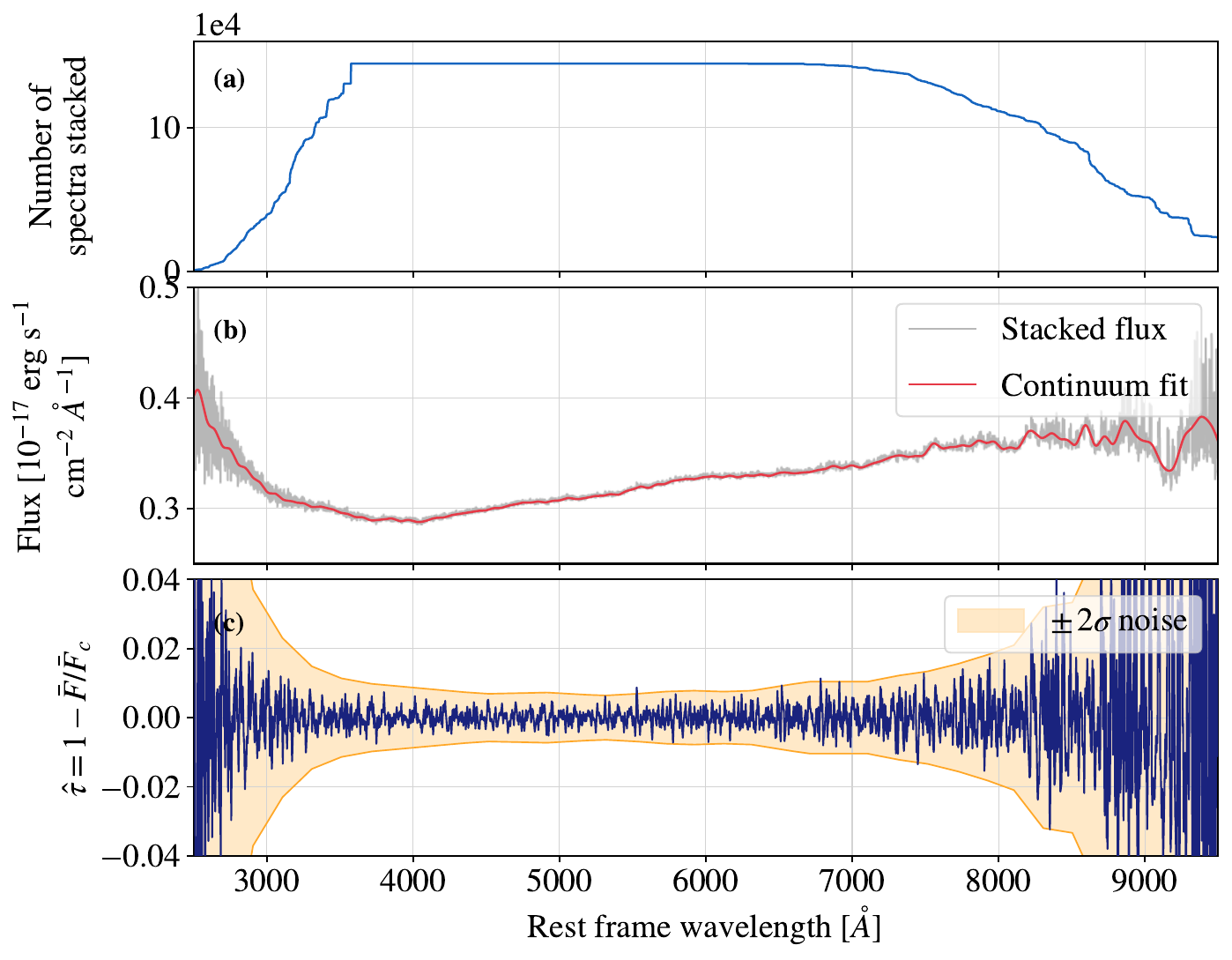}
        \caption{Stacked spectrum summary of DESI EDR--GAMA 
    source--absorber pairs. 
    \textit{Panel (a):} Number of spectra contributing to each rest-frame 
    wavelength bin. The plateau at $\sim\!3500$--$7000$\,\AA\ marks the region where nearly all pairs contribute; the roll-off at both ends reflects the finite DESI wavelength coverage combined with the spread of absorber redshifts. 
    \textit{Panel (b):} Inverse-variance weighted stacked flux (grey) and the  continuum estimate (red). The continuum captures the broad spectral shape of the source while leaving narrow features in the residual that could arise from a potential dark matter absorption line. 
    \textit{Panel (c):} Stacked optical depth $\hat{\tau} = 1 - \bar{F}/\bar{F}_c$ 
    (navy) with the $\pm 2\sigma$ noise envelope (orange). The noise is smallest at intermediate wavelengths, and grows toward the edges where fewer spectra 
    contribute. The optical depth is consistent with zero across the full range, with no feature exceeding the $2\sigma$ envelope. This allows us to place upper limits on the dark matter magnetic dipole moment.}
    \label{fig:stacking_summary}
    \end{figure}
The stacked optical depth is built using the following steps:
\begin{itemize}
    \item \textbf{Rest-frame transformation to a common grid.} For a given pair, the wavelength of the source spectrum is shifted to its absorber rest frame via $\lambda_\mathrm{obs}/(1+z_\mathrm{abs})$. The shifted flux and the associated inverse-variance weights are linearly interpolated onto the common grid running from $\lambda_\text{min} = \min(\lambda_\text{obs})/(1+z_\text{max})$ to $\lambda_\text{max} =\max(\lambda_\text{obs})/(1+z_\text{min})$, with a $0.8$\,\AA\ bin width matching the DESI pixel scale, where $z_\text{min}$ and $z_\text{max}$ are the minimum and maximum absorber redshifts.
 
    \item \textbf{Inverse-variance stacking.} The spectra are combined with inverse-variance weights to down-weight the low-quality pixels,
    \begin{equation}\label{eq:stack_flux}
        \bar{F}(\lambda) \;=\; 
        \frac{\sum_{i=1}^{N_\lambda}\, \sigma_i^{-2}(\lambda)\, F_i(\lambda)}
             {\sum_{i=1}^{N_\lambda}\, \sigma_i^{-2}(\lambda)} \,,
    \end{equation}
    where $F_i(\lambda)$ is the flux of the $i$-th spectrum and the sums run over $N_\lambda$ spectra contributing to each wavelength bin. 
 
    \item \textbf{Stacked optical depth.} The smooth continuum $\bar{F}_c$ is obtained from the low-frequency component of the Fourier transform of $\bar{F}$ and subtracted, which removes broad spectral features with widths $\gtrsim 100$\,\AA. Because the dark matter line is broadened only by the halo velocity dispersion, $\Delta\lambda/\lambda_* \sim \sigma_v/c \sim 10^{-3}$, its width of a few \AA\ lies well below this cutoff, and is preserved in the residual. The stacked optical depth then follows from the continuum-normalised flux,
    \begin{equation}\label{eq:tau_est}
        \hat{\tau}(\lambda) \;=\; 1 - \frac{\bar{F}(\lambda)}{\bar{F}_c(\lambda)}
        \qquad (\hat{\tau} \ll 1) \,,
    \end{equation}
    where the approximation holds in the optically thin limit expected for the weak dark matter transition.
    
    \item \textbf{Noise estimation.} The noise on the stacked optical depth is estimated empirically from the data, following the percentile method of~\cite{Wang2024}. We divide the spectrum into overlapping bins of width $400$\,\AA\ spaced every $200$\,\AA, and in each bin take the $16$th and $84$th percentiles of the $\hat{\tau}$ values to define the local $1\sigma$ noise as $\sigma_{\hat\tau} = \tfrac{1}{2}(q_{84} - q_{16})$. A smooth noise curve is then interpolated through these binned estimates. We verified that finer bins of $200$\,\AA\ spaced every $100$\,\AA\ yield a consistent noise level, though the resulting curve is less smooth owing to the smaller number of pixels per bin. 
\end{itemize}
The results are shown in Fig.~\ref{fig:stacking_summary}. The stacked optical depth is consistent with zero across the full wavelength range, with no feature exceeding the $2\sigma$ noise estimate. We translate this null result into an upper limit on $\mathcal{M}$ in Sec.~\ref{sec:constraints}.

\subsection{Deriving constraints}\label{sec:constraints}
 
We compare the observed stacked optical depth $\hat{\tau}(\lambda)$ to a theoretical prediction to constrain the dark matter parameters. For a given 
transition energy, we evaluate the optical depth from Eq.~\eqref{eq:tau} for each absorber using its redshift, mass, and impact parameter, and combine these into a predicted stack $\hat{\tau}_\text{th}(\lambda)$ with the same inverse-variance weights applied to the data, so that the predicted and observed stacks are constructed identically. For a given dark matter mass and transition energy, the predicted signal scales with the magnetic dipole moment as $\hat{\tau}_\text{th} \propto \mathcal{M}^2 \propto \sigma_\mathcal{M}$ (see Eq.~\eqref{eq:tau}). Requiring 
$\hat{\tau}_\text{th} < 2\,\sigma_{\hat{\tau}}$ yields an upper limit on $\sigma_\mathcal{M}$ at each $(\Delta E,\,m_\chi)$ parameter space.

\begin{figure}[!htbp]
        \centering
        \includegraphics[width=1.0\linewidth]{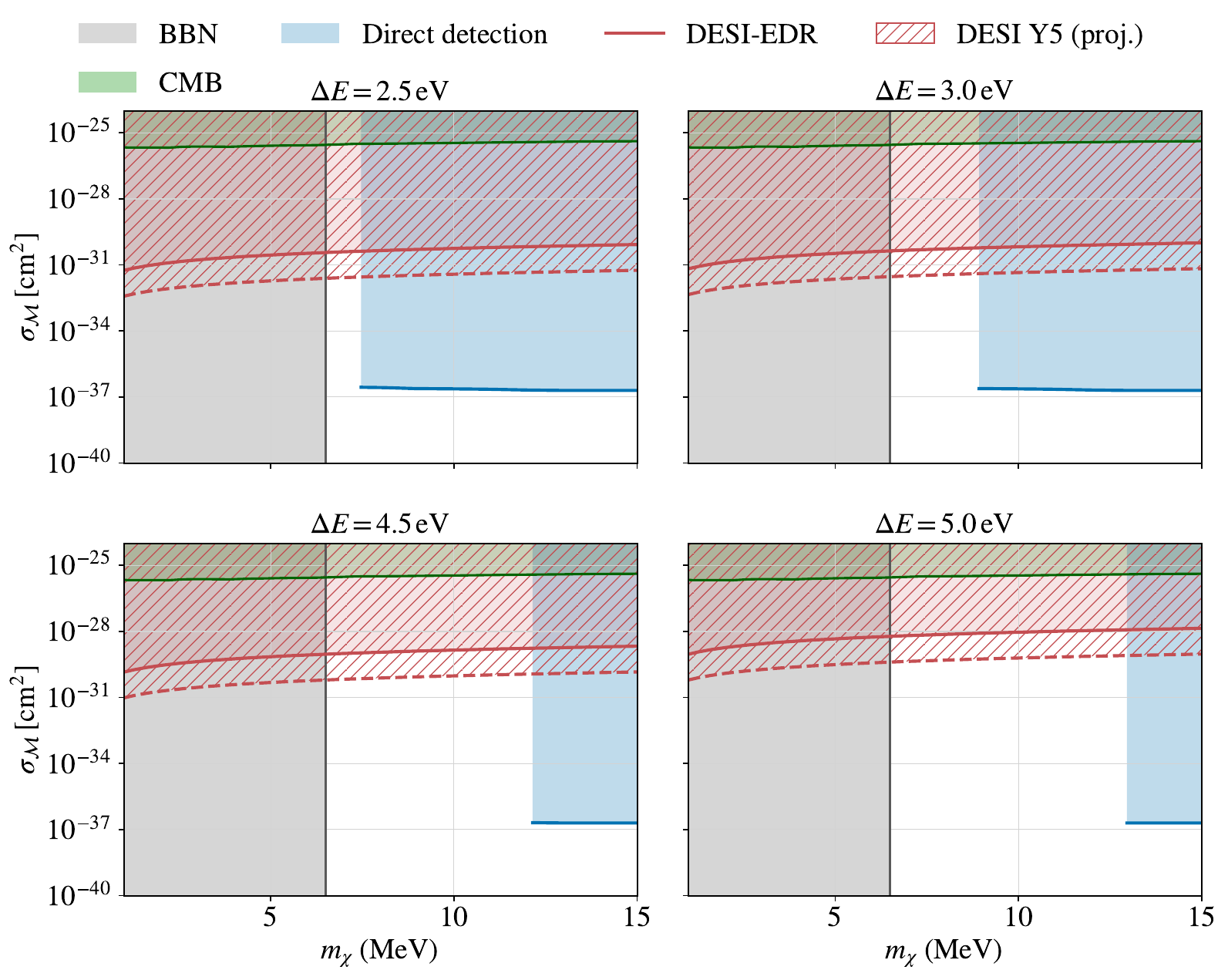}
        \caption{Upper limits on the magnetic dipole transition cross-section $\sigma_\mathcal{M}$ as a function of the dark matter mass $m_\chi$, shown for four values of the transition energy: $\Delta E = 2.5$, $3.0$, $4.5$, and $5.0$\,eV. The red solid lines show the $2\sigma$ exclusion from the DESI EDR--GAMA stacking analysis presented in this work, and the red hatched regions show the projected reach of the full DESI Year~5 survey, estimated using the scaling relation derived in Fig.~\ref{fig:forecast}. The vertical dashed line marks the thermal-relic mass bound, $m_\chi > 6.45$ MeV at 95$\%$ CL for an electromagnetically coupled Dirac fermion~\cite{An2022}. The scattering-only constraints on velocity-independent DM–proton scattering from CMB ~\cite{An2024} are shown as green shaded regions for elastic magnetic dipole dark matter. Elastic scattering is a good approximation at early times because the thermal energy of baryons is comparable to or greater than the eV-scale splitting.   Direct-detection limits (blue) rescale published DM–electron exclusion limits with $F_\mathrm{DM} = 1$~\cite{Damic2024, Sensei2026}. We assume that dark matter particles have a characteristic halo speed, $v_\chi \approx 300$ km/s, 
        can up-scatter, $m_\chi > 2\,(\Delta E + E_B)/v_\chi^2$, where $E_B \approx 1.2$ eV corresponds to the silicon bandgap, so that the elastic-limit rescaling holds. Exact limits from BBN, CMB, and direct detection for inelastic DM require a detailed calculation, which we leave for future work.} 
    \label{fig:exclusion}
    \end{figure}

\section{Results and forecasts}
    \label{sec:results}
    \subsection{Constraints from DESI EDR}
    \label{sec:edr_constraints}
    \begin{figure}[!htbp]
        \centering
        \includegraphics[width=0.95\linewidth]{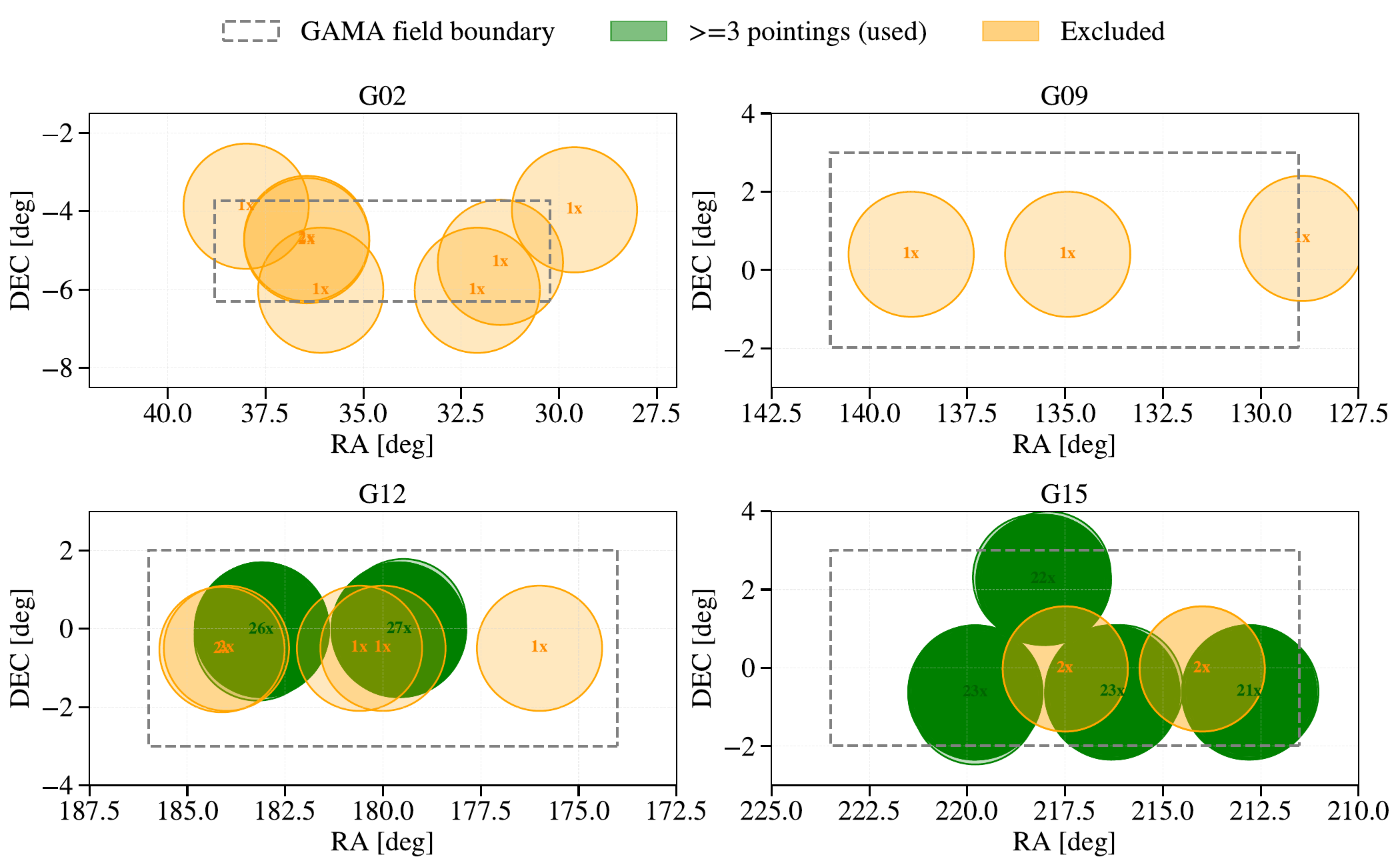}
        \includegraphics[width=0.65\linewidth]{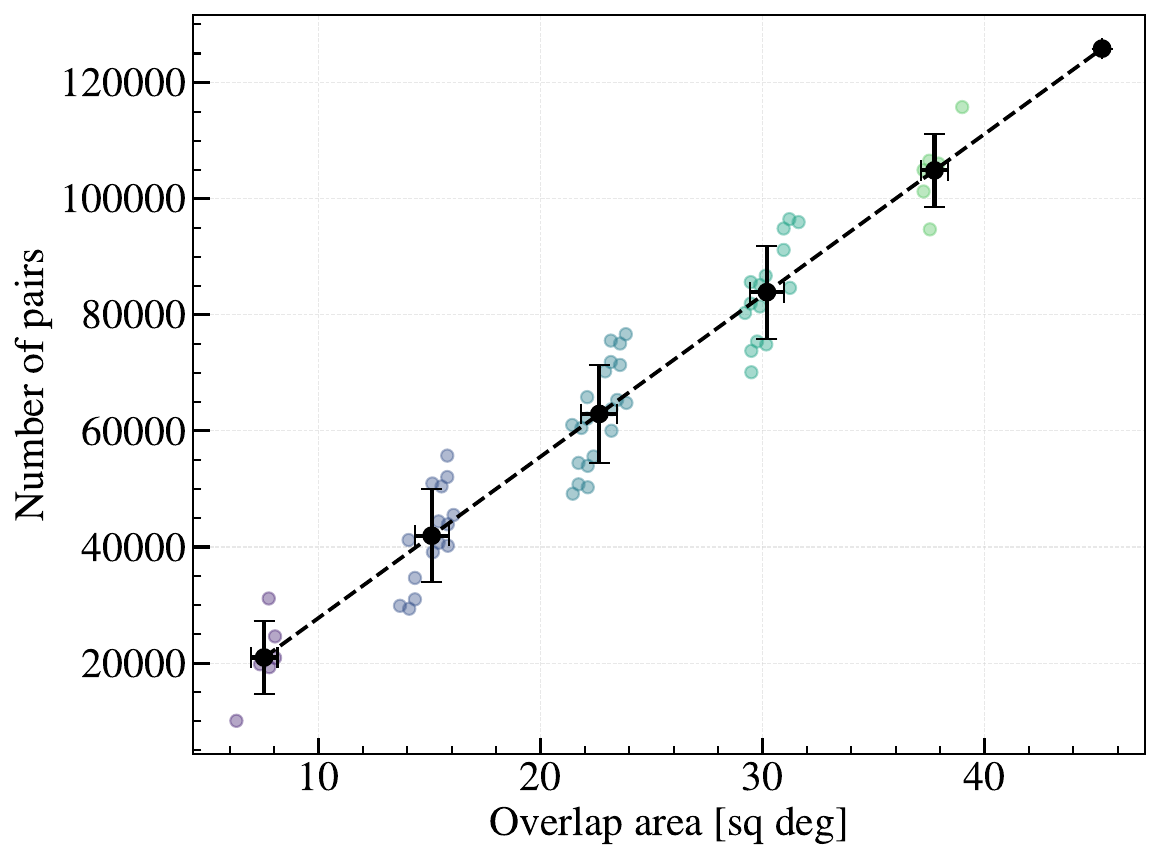}
        \caption{Identifying complete survey regions and calibrating the pair counts for sensitivity forecasts. 
    \textit{Top:} DESI EDR tile coverage over the four GAMA equatorial fields. Each circle represents a DESI tile, with the number of completed pointings 
    indicated. Tiles with $\geq 3$ pointings (green) provide a complete background source catalogue and are used to define six complete regions across the four fields; tiles with fewer pointings (orange) are excluded. Dashed rectangles mark the GAMA field boundaries. \textit{Bottom:} Number of source--absorber pairs as a function of the 
    overlap area, obtained by computing pairs within the six complete regions taken individually and in all possible combinations (coloured points). 
    The dashed line is a linear fit, confirming that the pair count scales proportionally with survey area. This empirical scaling is used to forecast 
    the sensitivity for larger spectroscopic surveys in 
    Sec.~\ref{sec:constraints}.}
    \label{fig:forecast_scaling}
    \end{figure}
    \begin{figure}[!htbp]
        \centering
        \includegraphics[width=0.8\linewidth]{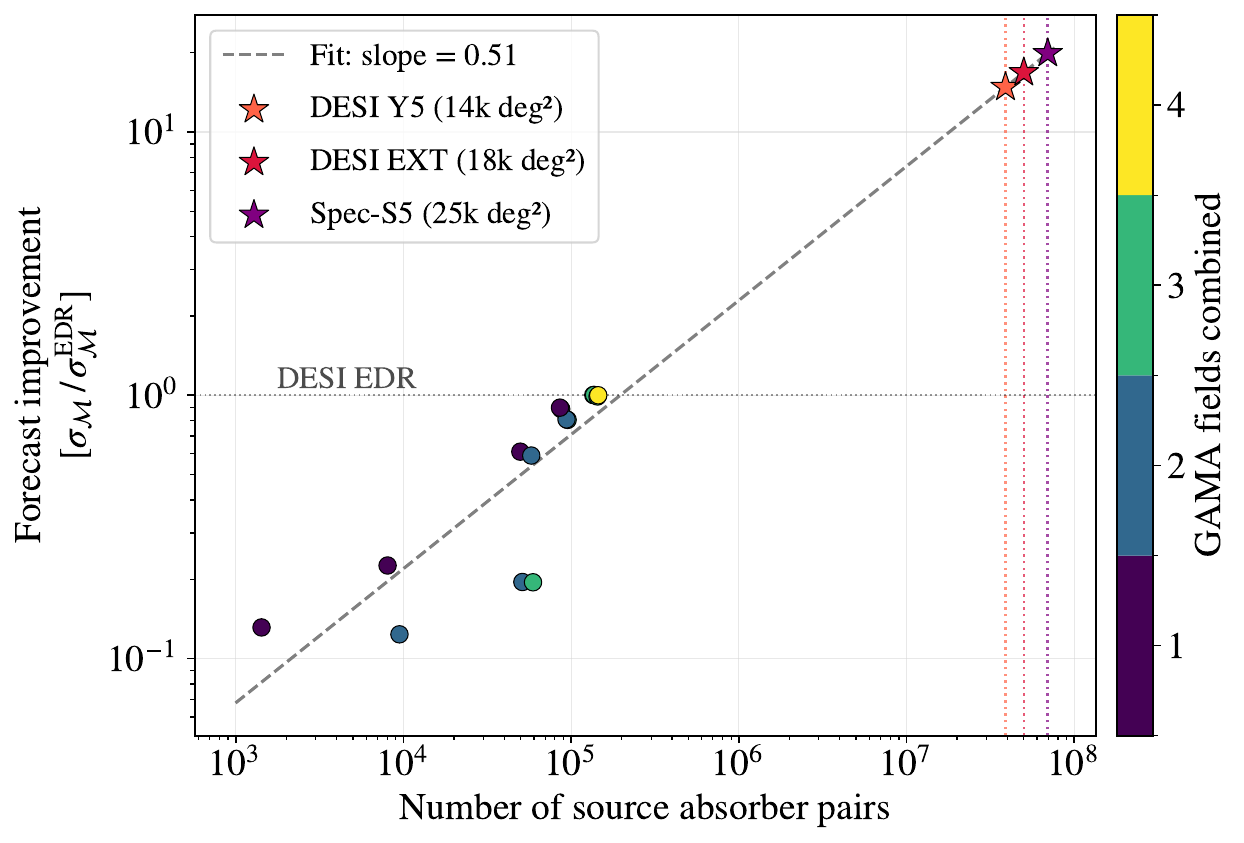}
        \caption{Forecast improvement on the upper bound on the magnetic dipole transition cross-section $\sigma_\mathcal{M}/\sigma_\mathcal{M}^\text{EDR}$ as a function of the number of source--absorber pairs, normalized to the DESI EDR baseline (horizontal dotted line). Filled circles show the empirical scaling obtained from the six complete DESI EDR regions (Fig.~\ref{fig:forecast_scaling}) taken individually and in combinations of up to four GAMA fields (colour bar). The dashed line is a power-law fit with slope $0.51$, consistent with the expected $\sqrt{N}$ scaling of the stacked signal-to-noise ratio. Star markers show the extrapolated sensitivity gain for the current and upcoming survey configurations: DESI Year~5, DESI Extended, and Spec-S5, with the corresponding pair counts indicated by vertical dotted lines. The full DESI Year~5 and Spec-S5 surveys are projected to improve constraints by more than an order of magnitude over the current EDR analysis.}
    \label{fig:forecast}
    \end{figure}

    Our main result is the set of $2\sigma$ upper limits on the magnetic dipole transition cross-section $\sigma_\mathcal{M}$ (Eq.~\ref{eq:sigmam}) 
    as a function of the dark matter mass $m_\chi$, shown in Fig.~\ref{fig:exclusion} for four transition energies $\Delta E = 2.5$, $3.0$, $4.5$, and $5.0$\,eV. The shape of the limits follows directly from how the predicted signal and the stack noise depend on the model parameters. For optical transitions $\Delta E \gg k_\mathrm{B}\,T_\mathrm{ex} \approx k_\mathrm{B}\,T_\mathrm{CMB}(1+z_\mathrm{abs})$, the predicted optical depth (see Eq.~\ref{eq:tau}) scales as $\tau \propto \mathcal{M}^2/m_\chi$, independent of $\Delta E$. At fixed $\Delta E$, the constraint therefore weakens with increasing $m_\chi$, since the lower number density of heavier DM particles reduces the absorption signal. The dependence on $\Delta E$ enters only through the stack noise. The stack noise $\sigma_{\hat\tau}$ rises steeply toward both ends of the DESI range (see third panel of Fig.~\ref{fig:stacking_summary}). The tightest constraints therefore arise at intermediate energies $\Delta E \sim 2$--$4$\,eV.
    
    At low masses, $m_\chi \lesssim 6$\,MeV, a dark matter candidate would be relativistic in the early Universe, changing the effective number of relativistic species affecting the expansion rate. If it also scatters electromagnetically with baryons and stays in kinetic equilibrium with the photon--baryon fluid. This modifies the CMB anisotropies, and observations exclude such candidates~\cite {Ho2013, Nollett2014, Escudero2019, Sabti2020, Sabti2021, An2022, An2024}. 
    The direct detection experiments using Skipper Charge-Coupled Devices probe dark matter electron scattering and set the leading limits~\cite{Damic2024, Sensei2026} at heavier dark matter masses. The up-scattering is endothermic, requiring the dark matter to carry enough kinetic energy to supply both the recoil and the transition energy. This requires a sufficiently heavy $m_\chi$, and the sensitivity falls away below the $\sim$ few MeV scale depending on the transition energy. Between these two regimes, for transition energies in $2.5$~-~$5$ eV, neither probe is effective. The spectroscopic stacking method developed here directly targets this intermediate region. The DESI EDR and the projected reach of the full DESI Year~5 survey (red hatched), derived from the scaling relations described in the next Sec.~\ref{sec:forecasts}, extends into parameter space inaccessible to both direct detection and cosmological probes.

\subsection{Forecasts for future surveys}\label{sec:forecasts}
    To forecast the sensitivity of spectroscopic surveys with larger footprints, we utilize two scaling relations: the improvement in the $\sigma_\mathcal{M}$ upper limit as a function of the number of pairs, and the number of source--absorber pairs as a function of survey area.
     
    For the pair counts, we use the six complete DESI EDR regions that overlap with the G12 and G15 GAMA fields (tiles with $\geq 3$ pointings, shown as green circles in the top panel of Fig.~\ref{fig:forecast_scaling}). The 
    completeness criterion ensures a uniform background source density, making the pair count a reliable proxy for survey area. The six regions have approximately equal overlap area, so taking all possible combinations results in six distinct area bins. The resulting pair count scales linearly with overlap area at a rate of $\sim\!2800\;\text{pairs\,deg}^{-2}$ 
    (shown in the bottom panel of Fig.~\ref{fig:forecast_scaling}).
     
    For the constraint scaling, we use all four GAMA fields (G02, G09, G12, G15), taken individually and in all possible combinations. Since the GAMA group sample falls predominantly within the DESI Bright Galaxy Survey(BGS) target class~\cite{Hahn2023}, we consider BGS galaxies to be the foreground absorbers in the forecast. The constraint improvement with respect to DESI-EDR, defined as $\sigma_\mathcal{M}/\sigma_\mathcal{M}^\text{EDR}$, follows a power law in the number of pairs with slope $0.51 \pm 0.02$ (see Fig.~\ref{fig:forecast}), 
    consistent with the $\sqrt{N}$ scaling expected when the noise on the individual spectra in a stack is uncorrelated.
     
    Combining these two scalings for the current and next generation surveys, we find sensitivity gains of $\sim\!15\times$ for DESI Year~5 ($14{,}000\;\text{deg}^2$)~\cite{DESI2016a}, $\sim\!17\times$ for DESI Extended ($18{,}000\;\text{deg}^2$)~\cite{DESIext2022}, and $\sim\!20\times$ for Spec-S5 
    ($25{,}000\;\text{deg}^2$)~\cite{SpecS2025}, relative to the DESI EDR baseline (star markers in Fig.~\ref{fig:forecast}). The projected exclusion region for DESI~Y5 is 
    shown in Fig.~\ref{fig:exclusion} (red hatched), where it extends well into parameter space currently unconstrained by either direct detection or 
    CMB.
    
\section{Conclusion}\label{sec:conclusion}
We summarize our main findings below:
\begin{itemize}
    \item We have introduced and demonstrated a new spectroscopic probe of electromagnetically interacting multi-state dark matter, based on the narrow absorption line produced when background light traverses a dark matter halo. We consider a two-state model where a magnetic dipole transition couples a ground state to a nearly degenerate excited state; for an eV-scale mass splitting, the resonant absorption falls in the optical band and becomes accessible to large spectroscopic surveys. 
    
    \item With 145{,}087 source--absorber pairs from the DESI Early Data Release and the GAMA group catalogue, we find a stacked optical depth consistent with zero across the optical range and place the first $2\sigma$ upper limits on the magnetic dipole transition cross-section $\sigma_\mathcal{M}$ over a few $\sim$ MeV mass range and for transition energies $\Delta E = 2.5$--$4.0$\,eV, in a parameter space inaccessible by CMB and direct detection experiments.

    \item The constraining power of our method will improve as the number of spectra increases. We forecast that our limits will strengthen by more than an order of magnitude for the full DESI Year~5 and future Stage 5 spectroscopic surveys.

    \item The absorption-line technique is not specific to the magnetic dipole model considered here and can be applied to any dark matter scenario that predicts new spectral signatures. As spectroscopic samples grow by orders of magnitude in the coming decade, stacked absorption-line searches offer a powerful and complementary route to uncover electromagnetic interactions of dark matter.
\end{itemize}



\acknowledgments
    We thank Rishi Khatri and Tuhin S. Roy for valuable feedback. This work was supported by the Department of Atomic Energy, Government of India, under Project Identification Number RTI-4012. The computations were carried out on the computing clusters at the Department of Theoretical Physics, TIFR, Mumbai. We also thank Ajay Salve and Kapil Ghadiali for their computational support. AG was supported in part by the Roman Project Infrastructure Team “Maximizing Cosmological Science with the Roman High Latitude Imaging Survey" (NASA contracts 80NM0018D0004-80NM0024F0012) and Department of Energy grant DE-SC0025993. GAMA is a joint European-Australasian project based around a spectroscopic campaign using the Anglo-Australian Telescope. The GAMA input catalogue is based on data taken from the Sloan Digital Sky Survey and the UKIRT Infrared Deep Sky Survey. Complementary imaging of the GAMA regions is being obtained by a number of independent survey programmes including GALEX MIS, VST KiDS, VISTA VIKING, WISE, Herschel-ATLAS, GMRT and ASKAP providing UV to radio coverage. GAMA is funded by the STFC (UK), the ARC (Australia), the AAO, and the participating institutions. The GAMA website is \url{http://www.gama-survey.org/}. This research used data obtained with the Dark Energy Spectroscopic Instrument (DESI). DESI construction and operations is managed by the Lawrence Berkeley National Laboratory. This material is based upon work supported by the U.S. Department of Energy, Office of Science, Office of High-Energy Physics, under Contract No. DE–AC02– 05CH11231, and by the National Energy Research Scientific Computing Center, a DOE Office of Science User Facility under the same contract. Additional support for DESI was provided by the U.S. National Science Foundation, Division of Astronomical Sciences under Contract No. AST-0950945 to the NSF’s National Optical-Infrared Astronomy Research Laboratory; the Science and Technology Facilities Council of the United Kingdom; the Gordon and Betty Moore Foundation; the Heising-Simons Foundation; the French Alternative Energies and Atomic Energy Commis; the National Council of Humanities, Science and Technology of Mexico (CONAHCYT); the Ministry of Science, Innovation and Universities of Spain (MICIU/AEI/10.13039/501100011033), and by the DESI Member Institutions: https://www.desi.lbl.gov/collaborating-institutions. The DESI collaboration is honored to be permitted to conduct scientific research on I’oligam Du’ag (Kitt Peak), a mountain with particular significance to the Tohono O’odham Nation. Any opinions, findings, and conclusions or recommendations expressed in this material are those of the author(s) and do not necessarily reflect the views of the U.S. National Science Foundation, the U.S. Department of Energy, or any of the listed funding agencies. We acknowledge the use of AI assistants (Google Gemini and Anthropic's Claude) for language editing and code assistance during the preparation of this manuscript. All scientific content, analysis, and conclusions are the authors' own and have been verified by the authors.

\newpage
\newpage

\bibliography{reference2.bib}
\bibliographystyle{unsrtads}
\end{document}